\documentclass[10pt,letterpaper]{article}
\usepackage[top=0.85in,left=1.5in,footskip=0.75in,marginparwidth=2in]{geometry}

\usepackage[utf8]{inputenc}

\usepackage{cite}

\usepackage{nameref,hyperref}

\usepackage{microtype}
\DisableLigatures[f]{encoding = *, family = * }

\raggedright
\usepackage{changepage}

\usepackage[aboveskip=1pt,labelfont=bf,labelsep=period,singlelinecheck=off]{caption}

\makeatletter
\renewcommand{\@biblabel}[1]{\quad#1.}
\makeatother

\usepackage{lastpage,fancyhdr,graphicx}
\usepackage{epstopdf}
\fancyheadoffset[L]{2.25in}
\fancyfootoffset[L]{2.25in}

\usepackage{color}

\definecolor{Gray}{gray}{.25}

\usepackage{graphicx}

\usepackage{sidecap}
\usepackage{algorithmic}
\usepackage{algorithm}
\usepackage{amsmath,amssymb,amsfonts}
\usepackage{subfigure}
\usepackage{wrapfig}
\usepackage[pscoord]{eso-pic}
\usepackage[fulladjust]{marginnote}
\reversemarginpar

\begin{document}
\vspace*{0.35in}

\begin{flushleft}
{\Large
\textbf\newline{Modelling DDoS Attacks in IoT Networks using Machine Learning}
}
\newline
\\
Pheeha Machaka\textsuperscript{1}, 
Olasupo Ajayi\textsuperscript{2,*},
Hloniphani Maluleke \textsuperscript{2},
Ferdinand Kahenga\textsuperscript{2},
Antoine Bagula\textsuperscript{2},
Kyandoghere Kyamakya\textsuperscript{3}

\bigskip

\bf{1} Department of Computer Science, University of South Africa, Pretoria, Gauteng, South Africa.
\\
\bf{2} ISAT Lab, Department of Computer Science, University of the Western Cape, Cape Town, South Africa.
\\
\bf{3} Institute for Smart Systems Technologies, Transportation Informatics Group, Alpen-Adria Universität Klagenfurt, Klagenfurt, Austria.

\bigskip
* olasupoajayi@gmail.com

\end{flushleft}

\begin{abstract} In current Internet-of-Things (IoT) deployments, a mix of traditional IP networking and IoT specific protocols, both relying on the TCP protocol, can be used to transport data from a source to a destination. Therefore, TCP-specific attacks, such as the 
Distributed Denial of Service (DDoS) using the TCP SYN attack, are one of the most plausible tools that attackers can use on Cyber-Physical Systems (CPS). This may be done by launching an attack from its IoT subsystem, here referred to as the ``CPS-IoT'', with potential propagation to the different servers located in both fog and the cloud infrastructures of the CPS. This study compares the effectiveness of supervised, unsupervised, semi-supervised machine learning algorithms, as well as statistical models for detecting DDoS attacks in CPS-IoT, particularly during data transmission to and from the physical space to the cyber space  via the Internet. The models considered are broadly grouped into three: i.) ML-based detection - Logistic Regression (LGR), K-Means, and Artificial Neural Networks (ANN) with two variants based on traffic slicing. We also looked into the effectiveness of semi-supervised hybrid learning models, which used unsupervised K-Means to label the data, then fed the output to a supervised learning model for attack detection. ii.) Statistic-based detection - Exponentially Weighted Moving Average (EWMA) and Linear Discriminant Analysis (LDA). iii.) Prediction algorithms - LGR, Kernel Ridge Regression (KRR) and Support Vector Regression (SVR). Results of simulations showed that the hybrid model was able to achieve 100\% accuracy with near zero false positives for all the ML models, while traffic slicing traffic helped improved detection time; the statistical models performed comparatively poorly, while the prediction models were able to achieve over 94\% attack prediction accuracy.
\end{abstract}

Anomaly detection; Distributed Denial of Service; Internet of Things; Machine Learning;  Regression analysis.


\section{INTRODUCTION}

The Internet of Things (IoT) provides a platform that allows objects to connect and communicate with one another using devices that can sense, identify and locate “things” in their surroundings, in order to better comprehend happenings in their environment. IoT devices are used for autonomous and intelligent tasks in residences, retail outlets, office buildings, transportation~\cite{ola1}, agriculture, healthcare~\cite{bag1,bag2}, and manufacturing  plants, among other places. The IoT market 
is growing at an exponential rate and is estimated to have grown to over 41 billion devices by 2027. Recently, the IoT has also expanded its reach beyond terrestrial networks by using drones~\cite{bag3} to complement the services delivered by semi-static IoT 
networks located on the ground~\cite{bag4,bag5}. The security of the complex network infrastructure resulting from the combination of terrestrial and airborne nodes, which use devices designed to operate in settings with limited resources (computing power, storage capacity, battery), is a challenging issue that requires incorporating security principles into different layers of the IoT protocol stack. For example, attacks such as Denial of Service (DoS) or Distributed DoS (DDoS) can be launched at the network, transport or application layers of the Internet stack, to easily compromise IoT devices \cite{2} when such devices run routing protocols that use these layers for the transport of sensor readings, as illustrated on Table~\ref{tab0}.

\begin{table}[H]
\centering
\caption{IoT Specific Protocols}
\label{tab0}
\setlength{\tabcolsep}{3pt}
{\small
\begin{tabular}{p{3.5cm} p{2cm} p{3cm} p{2cm} p{1.5cm}}
\hline
\textbf{Protocol} & \textbf{Underlying Protocol} &	\textbf{Architecture} &	\textbf{DDoS Prone} & \textbf{Ref.} \\ \hline
AMQP (RabbitMQ) & TCP & Publish/Subscribe & Yes & \cite{amqp} \\
CoAP & UDP & Request/Response & No* & \cite{coap}  \\ 
DDS & TCP & Publish/Subscribe & Yes & \cite{dds} \\ 
MQTT & TCP & Publish/Subscribe & Yes & \cite{mqtt} \\ 
XMPP & TCP & Both & Yes & \cite{xmpp} \\ \hline
\end{tabular}}
\end{table} 

Table~\ref{tab0}  shows some IoT, and/or message telemetry specific protocols, including Message Queuing Telemetry Transport (MQTT), Extensible Messaging and Presence Protocol (XMPP), Advanced Message Queuing Protocol (AMQP), and their corresponding underlying protocols. The table shows that protocols with publish/subscribe architecture rely on TCP protocol for data telemetry, and are thus susceptible to DDoS based TCP SYN attacks. It is important to note that though Constrained Application Protocol (CoAP) does not run on TCP, it is still vulnerable to DDoS attacks such as UDP Flood.

\begin{figure}[h]
    \centering
    \includegraphics[scale=0.5]{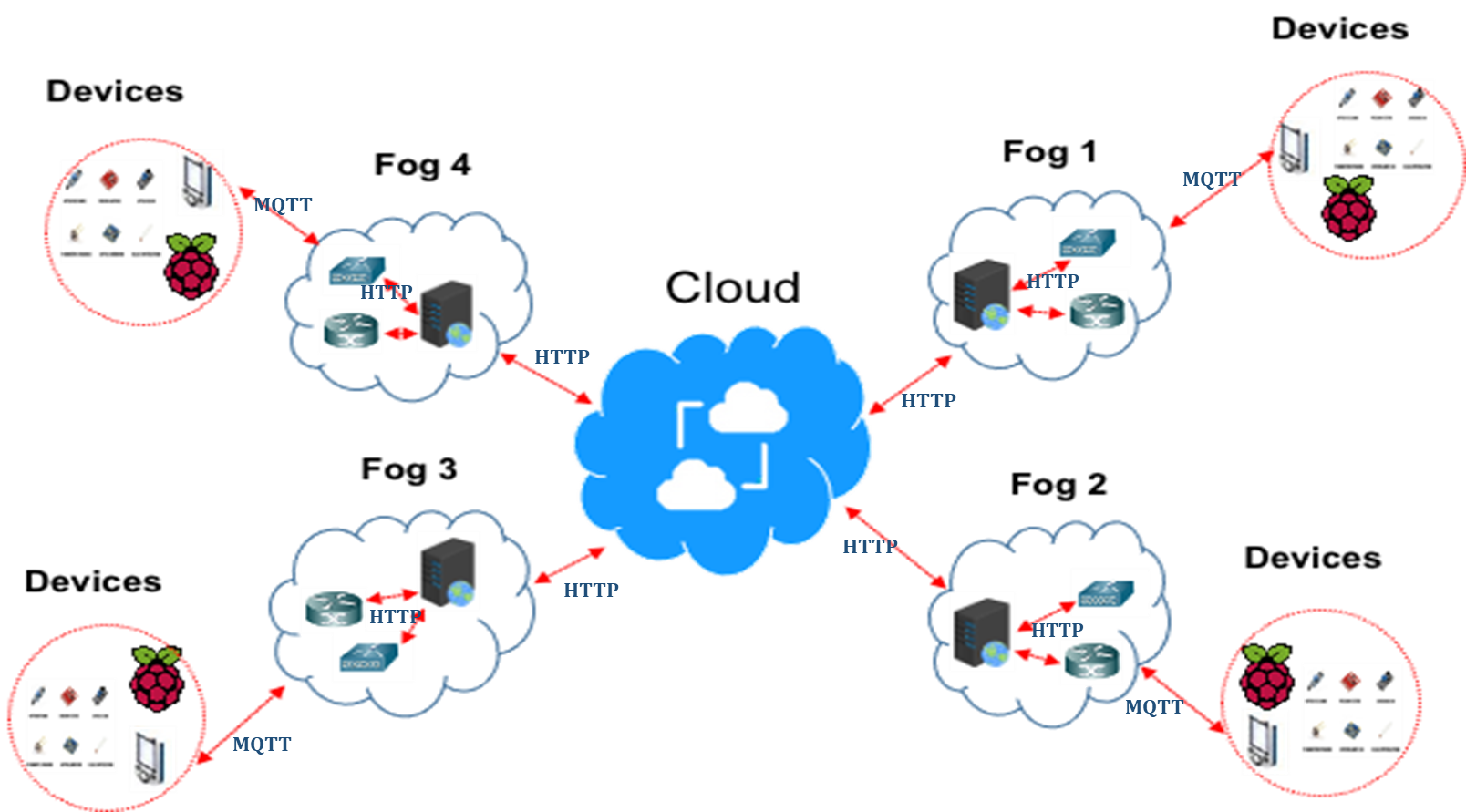}
    \caption{A Generic CPS-IoT Subsystem}
    \label{fig1}
\end{figure}

Cyber Physical Systems IoT subsystem (CPS-IoT) \cite{cpsiot}, such as that shown in Fig. \ref{fig1}, rely on a mix of traditional IP networks and IoT specific protocols to move data from devices (physical and virtual sensors, actuators, edge devices and gateways) to / from the Cloud. An IoT specific protocol, such as MQTT or AMQP, is used for message telemetry between device(s) and the Fog infrastructure, as shown in Fig \ref{fig1}; while an IP protocol, such as the HyperText Transfer Protocol (HTTP), is used between the Fog and Cloud infrastructures. While both protocols (MQTT and HTTP) belong to different stacks, they are both guided by the TCP protocol in transporting data from source to destination. Hence, DDoS attacks such as the TCP SYN can be plausible tools that attackers use to mislead the operation of CPS and potentially cause critical damages. 

Having shown through Fig. \ref{fig1} and Table \ref{tab0} that CPS-IoT data telemetry protocols mostly run on TCP/IP - HTTP (TCP port 80 or 8080) and MQTT (TCP port 1883 or 8883) \cite{mqtt} or AMQP (TCP port 5671 or 5672) \cite{amqp} - we now focus on modelling DDoS attacks on the underlying TCP/IP network layer in the rest of this paper. 

Due to the Internet's phenomenal development over the last few decades, attackers now have access to a growing number of vulnerable devices and often use the IoT subsystem of CPS (where these devices are located) to launch vicious attacks that can adversely affect the CPS as a whole. For instance, an attacker may use a large number of these susceptible devices to initiate an attack on a server located in a Fog close to the devices or in a Cloud infrastructure located far away. These attacks often have various modes of intensity, with attacks that are perpetrated with low intensities, often able to evade detection by current detection techniques. 

Through this research we explore the application of machine learning (ML) models, including classification and prediction, to model DDoS attacks, specifically SYN attacks in IP networks, such as those upon which CPS-IoT subsystems are built. A potential use case  of our research is in sensor virtualization in CPS-IoT systems. In this use case, virtual sensors are located in the Fog or Cloud infrastructure to enhance real sensors with the capability of differentiating  and classifying  incoming traffic into genuine or bogus traffic in real-time.  This discerning ability is a key requirement for the efficient operation of next generation CPS, where security would be paramount. The selection of the most efficient algorithms for the classification of the sensor data traffic and the prediction of future attacks on the CPS-IoT are other key requirements for ensuring the safe operation of CPS infrastructures. However, these processes are beyond the scope of this work. 

The specific contributions of this work include:

\begin{itemize}
    \item Comparison of the efficiency of supervised, unsupervised, semi-supervised ML models and statistical models in modelling DDoS attacks, in a bid to distinguish between safe and adversarial network traffic. 
    
    \item The development of a semi-supervised learning model, capable of auto-labelling traffic and using the labelled traffic to accurately identify malicious traffic. This is achieved by hybridizing supervised and unsupervised machine learning models.
    
    \item Determining the impact, if any, of splitting network traffic into window sizes versus using the entire traffic stream in detecting malicious attacks.
    
    \item Exploring the effectiveness of regression models in predicting potential DDoS attacks, in a bid to move the safety of IP networks from reactive to proactive. 
\end{itemize}  

The rest of the paper is structured as follows, related literature are reviewed in Section \ref{s2}, while our research methodology is presented in Section \ref{s3}. Section \ref{s4} gives details of our implementation process and obtained results, while Section \ref{s6} concludes the paper and gives insights into potential future research directions.

\section{Literature Review} \label{s2}

The first DDoS assault on the public Internet happened in August 1999 \cite{3}. In February 2000, a year after the initial event, several commercial websites, including Yahoo, CNN, and eBay, saw their first DDoS attacks. A high number of requests overloaded these websites, forcing their services to go offline which  resulted  in considerable financial losses. The July 4 2009 cyber-attacks are well-known examples of DDoS attack, where prominent government, news media, and financial websites were targeted in a series of cyber-attacks across South Korea and the United States \cite{4}. These attacks caused service interruptions and the loss of millions of dollars each hour while companies were fighting to restore their Internet services.

Parallel to the attackers’ mode of operation becoming more sophisticated and with global reach, researchers have been investigating and developing defence mechanisms against DDoS attack. There are various types of defence mechanisms that have been developed so far and are broadly classified as signature-based or anomaly-based detection systems.

Signature-based detection systems try to create a collection of templates (signatures or rules) that may be used to determine if a particular network traffic pattern represents an intrusion. If the attack falls into one of the attack classes specified in the database, it can be effectively detected or recognized. As a result, signature-based systems are capable of detecting known intrusions (patterns) with high accuracy and a low number of false positives. However, they perform poorly in detecting novel / unknown attacks or variations of existing attacks \cite{5}. 
 
The limitations of signature-based intrusion detection motivated the development of anomaly-based detection systems. Anomaly-based intrusion detection systems (ABIDS) are concerned with detecting occurrences that appear to be out of the ordinary in terms of system behaviour. When a divergence from regular traffic behaviour is noticed, an attack is reported. ABIDS seek to distinguish between regular network activity and abnormal network activity. This is accomplished by developing a "normal system profile" based on previous data, which serves as a baseline. They then monitor for activities that are substantially different from this baseline. ABIDS are able to detect variations in "normal" network traffic patterns, such as sudden spikes, as possible DDoS attacks. Statistical characteristics can also be used, such that a sudden shift in variance or standard deviation can be flagged as potential attack(s) \cite{6}. 

In works relating to security of IoT, the authors in \cite{khan} surveyed IoT related security challenges and potential solutions for attacks such as DoS.  In \cite{opp}, a multi-layer defence mechanism was proposed for securing IoT data transmission between sensor nodes and gateways in rural communities with limited Internet access. In \cite{cpsiot} mathematical epidemiology was employed to monitor the safety and dependability of CPS-IoT systems. The authors built an orchestration model to monitor inbound traffic and make necessary adjustments to the actuation process. Furthermore, a protocol for node reconfiguration of the IoT sensor network during periods of attack was also proposed. In \cite{suv} IoT related attacks were split into four major categories, viz.: physical attacks, including tempering, radio frequency interference, and DoS at the perception layer; network attacks, including man in the middle attack, replay attack and network layer [D]DoS attacks; software attacks, including malwares and viruses; and data attacks, including data breach and unauthorized access. 
 
Detection methods based on various models and theories have been developed in the DDoS attack detection research community. The three key technologies that form the basis of the majority of today's detection techniques are machine learning (ML), information theory, and statistical models \cite{7}. Artificial Neural Networks (ANN), support vector machine (SVM), and other ML techniques in cybersecurity are helpful for decision making analysis \cite{8}. The paragraphs that follow highlight some of the related work in application of ML to DDoS attack detection. 
 
Ali \textit{et al.} \cite{9} developed an ANN-based ML strategy for detecting DDoS attacks. The backpropagation strategies employed by the ANN were Bayesian Regularization (BR) and Scaled Conjugate Gradient (SCG). The approach effectively detects DDoS attacks with an accuracy of up to 99.6\% using BR and 97.7\% using SCG backpropagation algorithms. In \cite{10} the authors created a system for detecting numerous large-scale IoT attacks in sequential order. They proposed using different specified classifiers for each attack type instead of a single classifier. For experimentation, they presented a single-layered ANN using publicly available datasets. They used a series of ANN models to detect specific assault types and were able to achieve 99\% accuracy by using the sigmoid function. In order to detect DDoS attacks, the authors in \cite{11} proposed combining feature selection with an ANN MLP (multilayer perceptron) model. This strategy was used to choose the best features during the training phase, and they created a feedback system to reconstruct the detector when significant detection faults were detected dynamically. With a 98\% accuracy rate, the proposed methodology proved effective. 
 
Ref. \cite{12} proposed a supervised learning anomaly detection model that combines a Radial Basis Function (RBF) kernel with a C-support optimizer (c-SVM) to differentiate between benign and malicious traffic data. With Blackhole and Sinkhole attacks, the model was 100\% accurate, whereas with other attack types, it was 81\% accurate. However, the researchers did not compare the outcomes of their experiment with those of other ML models. Chaudhary \textit{et al.} \cite{13} also suggested a ML technique for detecting DDoS assaults that involved filtering crucial network packet parameters such as packet size and interval size. SVM, Random Forest, Decision Tree, and Logistic Regression were used and Random forest surpassed the other models with a DDoS attack detection accuracy of 99.17\%.  In \cite{14} a method for detecting DDoS attacks using the SVM classifier was proposed. The SVM classifier had a 0.8\% false alarm rate and a classification accuracy of 95.11\%. In \cite{15}, the authors used flow features of network traffic, such as packet size, packet interval, protocol, bandwidth, and destination IP, to construct a model to detect DDoS attacks. They used SVM, K-Nearest Neighbour (KNN), Random Forest, Decision Tree, and ANN in their models. The results of the experiment showed that Random Forest and ANN have 99\% accuracy in detecting malicious traffic. 
 
For detecting DDoS attacks in Software Defined Networks (SDN), \cite{16} employed SVM, KNN, ANN, and Naive Bayes. Initially, the authors specified twelve features, but the algorithms chose a subset of these features based on threshold values. The algorithms analyzed flow traffic data and detected DDoS with 98.3 \% accuracy. In a similar study done in \cite{17}, ANN, SVM, Logistic Regression, KNN, Gaussian Naive Bayes, Bernoulli Naive Bayes, Multinomial Naive Bayes, Decision Tree (entropy-gini), and Random Forest algorithms were all investigated for DDoS attack detection. They looked at data from twelve different aspects and discovered that only a small fraction of them, such as cumulative count and descriptive statistics, was enough to detect a DDoS attack. In their tests, they discovered that the SVM algorithm had the highest accuracy rate of 99.7\%. Ref. \cite{18} used 23 traffic flow features to look into DDoS attack detection in SDN. They employed the Neighbourhood Component Analysis (NCA) to determine the most important flow data characteristics for the pre-processing and feature selection stage. Following that, they classified DDoS attacks using the KNN, Decision Tree (DT), ANN, and SVM algorithms. They discovered 14 features to be important in their findings, and the DT algorithm was able to attain 100\% detection accuracy. Authors in \cite{18b} used KNN, SVM, decision tree (DT), naïve Bayes (NB), Random Forest (RF), ANN, and logistic regression (LGR) algorithms to explore the detection of DDoS attacks in IoT networks. Their research looked into the effectiveness of algorithms for binary and multi-class classification. They also tested the algorithms' performance against a weighted and non-weighted Bot-IoT dataset. For non-weighted datasets, their testing revealed that the RF algorithm has a 99\% accuracy. The ANN performed better in binary classification accuracy on weighted datasets. KNN, on the other hand, surpassed other ML algorithms in multi-class classification, with an accuracy of 99\%, which is 4\% higher than RF.

The accurate and timely detection of DDoS attacks remains a priority for researchers in the field of cybersecurity, however, attackers keep modifying and developing new attacks in order to evade detection techniques. In this research study we distinguish between normal and DDoS attack network traffic and compare the performance of supervised, unsupervised, and semi-supervised machine learning techniques. Additionally, the efficacy of two approaches for forecasting possible DDoS attacks was investigated. In the section that follows, we will provide a detailed account of the methodological approach followed in this study.

\section{Methodology} \label{s3}
Fig. \ref{fig2} gives an overview of the proposed system with major components being data pre-processing, supervised learning, semi-supervised learning, unsupervised learning and prediction. Each of these components described as follows:

\begin{figure}[h]
    \centering
    \includegraphics[scale=0.45]{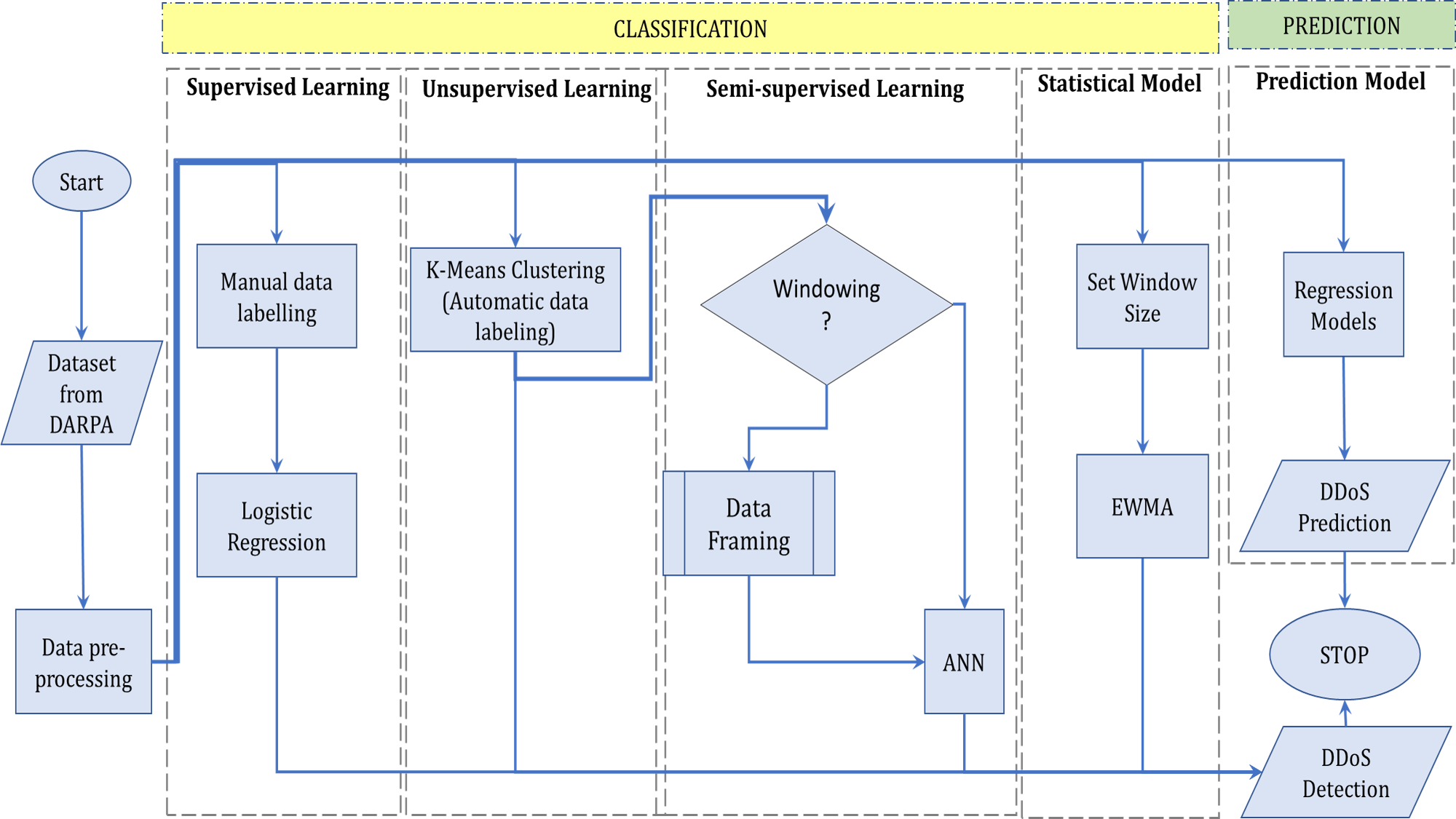}
    \caption{General System Architecture}
    \label{fig2}
\end{figure}

\subsection{Data Pre-processing \& Labelling}
For this work, we used the DAPRA IDS evaluation dataset \cite{19}, which was prepared by the MIT Lincoln Laboratory under DARPA and AFRL sponsorship. We used this dataset because we had earlier inferred that IoT systems have an underlying IP network upon which they run, hence still vulnerable to classic IP attacks such as DDoS. The tcpdump format was used, wherein all network activities, including the whole payload of each packet, were recorded and supplied for assessment. The data used in these evaluations were sniffed network traffic, Solaris BSM audit data, and Windows NT audit data.  Finally, the test network was made up of a mix of real and simulated machines, with the real and simulated machines artificially generating background traffic while the attacks were carried out on the real machines.

\begin{figure}[h]
    \centering
    \includegraphics[scale=0.45]{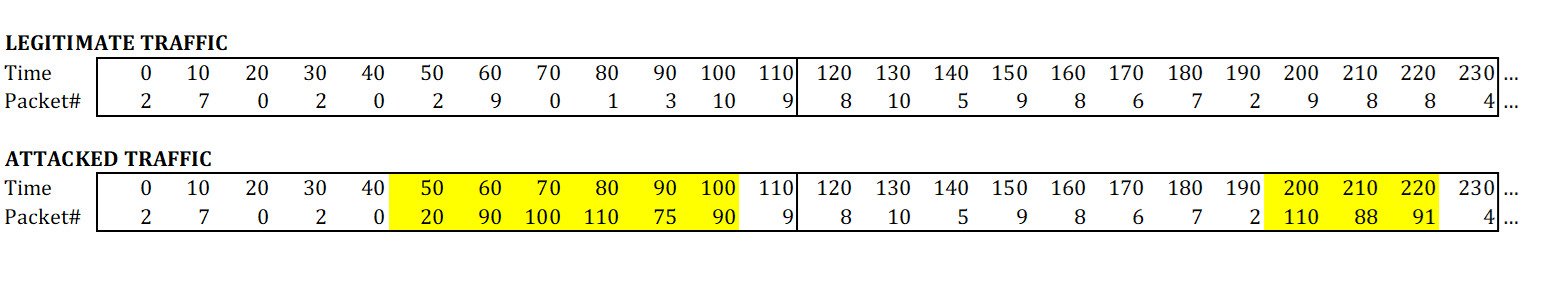}
    \caption{Data labels for Legitimate and Malicious Traffic}
    \label{fig3}
\end{figure}

We processed the raw dataset by writing a Python script to count the number of network packets that arrived at a given host per 10 second interval. We used this as the base line each 10 second block as zero (0), corresponding to no DDoS attack \cite{20}. We then introduced malicious attacks to the dataset by manually increasing the number of packets arriving in randomly selected intervals. We labelled these as one (1), implying DDoS attack. Fig. \ref{fig3} gives a visual illustration of the process and shows packet count for the first 2 minutes (120 seconds) of traffic flow. The top image represents legitimate (normal) traffic, while the bottom image depicts introduction of DDoS traffic (highlighted in yellow).

\subsection{Supervised Learning}
This component is labelled "Supervised Learning" in Fig. \ref{fig2} and it involved applying supervised ML on manually labelled data. Supervised learning is a class of machine learning (ML) wherein an ML model is trained using pre-labelled data, which serve as “examples” for the ML model. Once the model has been trained, it can then be exposed to new (test) data for classification or prediction. In our system, we considered Logistic Regression (LGR) and Artificial Neural Network (ANN) models.

\subsubsection{Data Framing}
Data framing was done for ANN only and three variants were considered. In the first, data framing was not considered, and this served as the baseline; while in the second, the dataset was split into "frames” of size 12, corresponding to 120 seconds of traffic flow (at 10 seconds interval). In the third, the standard deviation of values in the frame was calculated and appended to the frame, thus increasing the frame size to 13. The data frames were then fed to the ANN model. The data framing process is summarized with the pseudocode in algorithm \ref{alg1}.

\begin{algorithm}
\caption{Data Framing Algorithm}
\label{alg1}
\begin{itemize}
    \item Divide the entire dataset into data blocks of 120 seconds.
    
    \item For each 120 second data block in the dataset:
    
    \begin{enumerate}
        \item Create a 3 by 4 dataframe as follows: 
        
        \begin{itemize}
            \item Set t = 0
            \item For row = 1 to 4
            \item[a.] $col1 = Packet-Count (t), t += 10$
            \item[b.] $col2 = Packet-Count (t), t += 10$
            \item[c.] $col3 = Packet-Count (t), t += 10$
        \end{itemize}
        
        \item Calculate the stand deviation $(\sigma)$ for the data block. //for option 1 only
    \end{enumerate}

\end{itemize}

\end{algorithm}

\begin{figure}[h]
    \centering
    \includegraphics[scale=0.7]{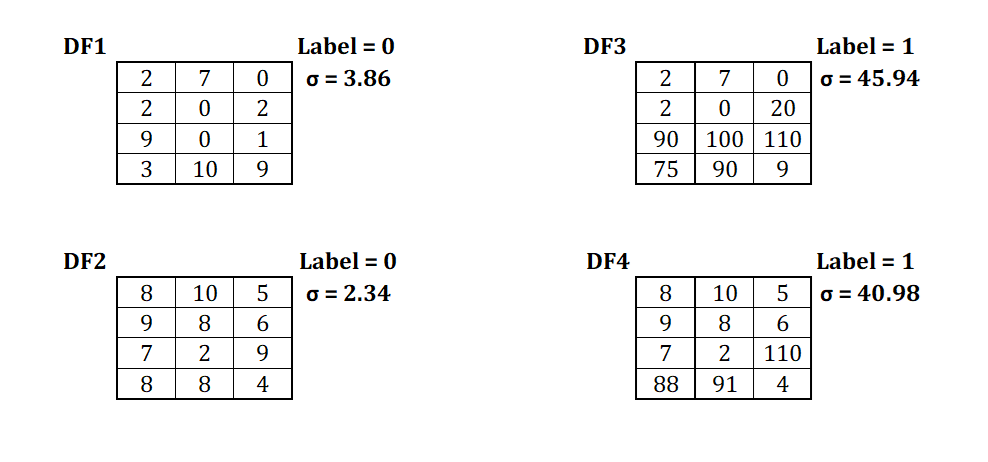}
    \caption{Sample Dataframes}
    \label{fig4}
\end{figure}

As an illustration, by running algorithm \ref{alg1} on the sample data presented in Fig. \ref{fig3} we end up with four data frames, which are shown in Fig. \ref{fig4}. In the figure, DF1 and DF2 are data frames representing legitimate traffic (top image in Fig. \ref{fig3}) and are labelled 0 which means there is no DDoS attack. DF3 and DF4 both represent traffic flows with malicious attacks, hence labelled 1. 

\begin{figure}[!htb]
    \centering
    \includegraphics[scale=0.65]{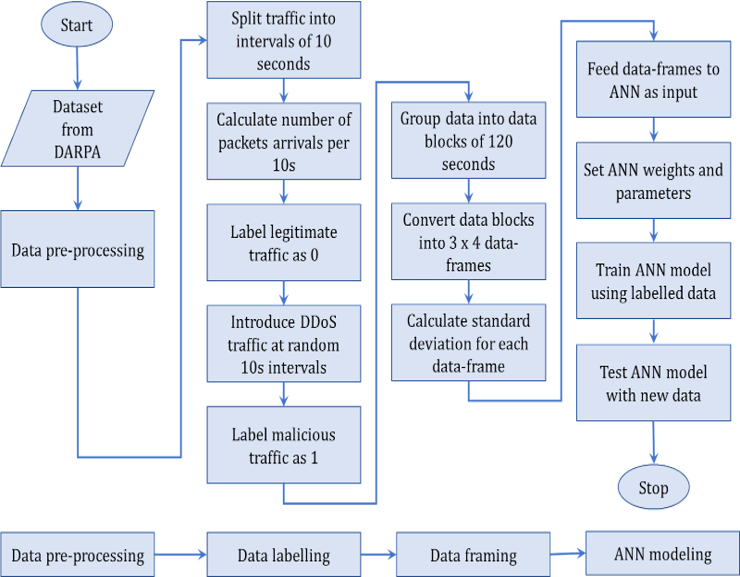}
    \caption{ANN Supervised Learning Process}
    \label{fig5}
\end{figure}

For each data frame, the standard deviation $(\sigma)$ is calculated. This standard deviation is used to further verify the probability that a malicious attacks has occurred. Within a data frame, if the data points are far from the mean, then the deviation of values within the frame would be higher, which implies that an attack occurred in that data frame. The opposite holds true for data points that are closer as their deviation from the mean would be smaller. This can be interpreted as an absence of attack(s). Finally, in instances where all 12 entries in a data frame are high (full DDoS attacks), the standard deviation value from the mean would be small. To distinguish between this full attack situation and a safe situation, a threshold value is used. If the calculated $\sigma$ is greater than this threshold value, then the frame is classified as being under attack.

\subsubsection{Machine Learning Models}

As mentioned above, both LGR and ANN were considered in this work. For LGR, an 80:20 split was used for training and test data, using the One-over-rest (OvR) training scheme and linear memory bfgs solver. ANN on the other hand is a model that mimics the human nervous system and has been used to solve numerous problems in Computer science \cite{21}, such as image recognition \cite{22}  and network intrusion detection \cite{23}. In this work, we used it to model DDoS attacks in IP networks and we considered a 3 layered ANN architecture. At the input layer we had 12 or 13 nodes ($\sigma$ included), the hidden layer had 6 nodes, while the output layer had 1 node. The data frames obtained in the previous subsection were fed in, with the corresponding standard deviation value used as the 13th node. The ReLu activation was used for the input and hidden layers, while Sigmoid activation was used at the output layer. The processes involved in our ANN supervised learning component are depicted in Fig. \ref{fig5}. 

\subsection{Unsupervised Learning}
As an alternative to manually labelling the dataset, we considered the K-means clustering technique for automatic labelling. K-Means is a centroid based clustering algorithm that determines cluster membership based on the proximity of data points to a centre point (centroid) \cite{24}. It  has been used in numerous classification problems including network classification \cite{25}, intrusion detection \cite{26} and in trans-continental networking \cite{27}.

In IP networks security, millions of packets often traverse the network per unit time and need to be classified (labelled) as either legitimate or malicious traffic. Manually doing this would be slow and laborious in such cases, hence the use of an automatic classifier is desirable, in our case K-Means. In our work, traffic flow fall into one of two categories (legitimate or malicious), thus,  k value is set to 2. 

\subsection{Semi-Supervised Learning}
Our semi-supervised learning component, which is labelled "Semi-Supervised Learning" in Fig. \ref{fig2}, is similar to the supervised learning described earlier. The major difference is that rather than using manually labelled data as input to the ML models, we fed the output of the unsupervised learning (K-Means clustering) into the models. In essence, K-Means is used to automatically label (classify) the data, which is then used to train the supervised model. By doing this we create a hybrid of a supervised and unsupervised model, or in this context, a semi-supervised model. The output of this model is then compared to the two other models (supervised and unsupervised).

\subsection{Statistical Models}
For completeness, we performed data classification using a classic statistical model - the Exponentially Weighted Moving Average (EWMA) \cite{roberts}. By placing more emphasis on recent data points than on older ones, EWMA is able to detect anomalies in observed data quickly. It has been applied in analysing and forecasting stock market, weather, and network traffic \cite{ye, yu}. 

In applying EWMA, we sought to detect anomalies within blocks of data. An anomaly would be a disruption from the "norm" (normal traffic flow). Such anomalies are considered as attacks. We set a window size of 12, corresponding to 120 seconds, then measured the deviations from the average traffic count in each window. The steps for calculating EWMA are well documented in literature, however, a concise summary of our application is given in algorithm \ref{alg2}  

\vspace{1mm}
\begin{algorithm}[H]
\caption{EWMA Algorithm}
\label{alg2}
\begin{itemize}
    \item Set a window size of 12 (data blocks of 120 seconds).
    \item Set mean, std, thresholdUp, thresholdDw to 0
    \item For each window:
    
    \begin{enumerate}
        \item mean += calculate the moving average.
        \item std += calculate the moving standard deviation.
        \item $thresholdUp = mean + std$
        \item $thresholdDw = mean - std$
        \item Slide the window by 1 (10 seconds)
        \end{enumerate}
        
    \item For each data point (d) in the dataset:
    \begin{enumerate}
        \item $if d > thresholdUp$ or $d < thresholdDw$: attack = True.
        \item Else attack = False
    \end{enumerate}
\end{itemize}
\end{algorithm}

Linear Discriminant Analysis (LDA) is a statistical model used for feature reduction and distinguishing between data entries in a dataset. In this work we are only concerned with its application in data classification abilities, specifically binary classification of data traffic into normal or attack. LDA has been used in similar applications domains such as in \cite{ldaIndia, noel}, where it was used for anomaly detection in network traffic.

This classification (discrimination) process of LDA is achieved using Bayes' theorem. The steps of the 2-dimensional (binary) LDA classification are well documented in literature, having initially being proposed in the early 1930s by Fisher \cite{fishLDA}. For brevity the steps are repeated in this work but refer interested readers to the work of Fisher for details.     

\subsection{Prediction}
Having successfully classified and distinguished between legitimate and malicious attacks, the next logical step might be to predict the possible occurrence of such attacks. This would help the network administrator put preventive measures in place to mitigate them, essentially changing the defense strategy from reactive to proactive. This is highlighted in green in Fig. \ref{fig2}. Three regression models were considered in this work for prediction, the Logistic Regression (LGR), Kernel Ridge Regression (KRR) and Support Vector Regression (SVR).

LGR models the probability that an event would occur, called the dependent variable based on one or more independent variable(s). We chose LGR because it is well suited for finding binary output probabilities, i.e. True or False (1 or 0) and it does not require a linear relationship between the dependent and independent variables. In applying it to our work we used it to determine if there would be an attack at a given time in future. The output variable was ‘time’, while the independent variables were  traffic ‘count' and ‘status’ (i.e. legitimate (0) or attack (1)). SVR  is a version of support vector machines proposed in \cite{28} that tries to minimize the predictor coefficient to a value less than or equal to a set threshold. We used the Guassian radial basis function (rbf) as the kernel and grid search to find optimal parameters. 
Similar to SVR, KRR also uses the kernel trick (rbf) but it uses a ridge as loss function instead of epsilon used by SVR. KRR combines kernel trick with least square regression and has been shown to be faster than SVR \cite{krr1, krr2}. We chose KRR because of its similarity with SVR and because there are large variances (deviations from mean) between data points in the dataset. These variances become more pronounced when the packet count in legitimate traffic flow (labelled 0) are compared with those of attacks (labelled 1).

\section{Implementation} \label{s4}
For this work, implementation was carried out on Google Colab, with a Python 3 Google Compute module, configured with 12GB of RAM, 2.3GHz 2 Core Intel Xeon CPU and GPU hardware accelerators. Keras and Sci-Kit learn were used for machine learning; Smote was used for data balancing; Pandas, NumPy were used for data manipulation, while matplotlib was used for data visualization. Finally, an 80:20 split was used for training and testing data for the supervised learning algorithms. 

\subsection{Metrics}
Six metrics were used to compare the performance of the models considered, these are false positive, false negative, average execution time, accuracy, $R^2$, and Root Mean Square Error (RMSE). The first 4 are specific to classification models, while the last 3 (accuracy inclusive) are for the regression models.  False positive is the number of malicious traffic that were misclassified as legitimate traffic. This is important as it measures how well the model is able to detect attacks. False negative measures the number of legitimate traffic that were misclassified as attacks.  We define average execution time as how long it takes the model to classify traffic. 
Precision and Recall are metrics that are well used in literature, but are not considered in this work because they both measure how “correct” a model is. Though these are important, for us, the inaccuracy of a model is more important, as undetected attacks can be catastrophic. Accuracy is a measure of the model’s classification (or regression) performance, i.e. number of traffic that were correctly identified by the model. $R^2$ is used to benchmark the performance of a model against a baseline, while RSME is the square root of the mean squared difference between predicted and actual values.

\subsection{Supervised Learning}
Table \ref{tab1} summarizes the results of the supervised learning models. 

\vspace{1mm}
\begin{table*}[!htb]
\centering
\caption{Anomaly Detection Using Supervised Learning Models}
\label{tab1}
{\footnotesize
\begin{tabular}{p{3.5cm} p{2cm} p{2.7cm} p{2.9cm} }
\hline
\textbf{Model} & \textbf{Accuracy} &	\textbf{False Positives} &	\textbf{False Negatives} \\ \hline 
LGR	& 99.192 & 0 & 1.6215 \\ 
ANN & 99.414 & 0 & 0.6695 \\ 
ANN+Dataframing & 98.842 & 2.1805 & 0.1295 \\ 
ANN+Dataframing+ SD. & 99.405 & 0.9565 & 0.0965 \\ 
\\ \hline
\end{tabular}}
\end{table*}

From the table, the pure ANN model resulted in the highest accuracy, followed by ANN + Data framing + SD. Logistic Regression (LGR) also had high accuracy value but resulted in the highest number of false negatives, meaning that LGR wrongly classified more legitimate traffic as attacks. Conversely, data framing adversely affected the false positive rate, resulting in about 2\% of bogus traffic (DDoS attacks) being misclassified as safe. Among the three ANN models considered, the variant without data framing only slightly edged out the variant with data framing and standard deviation (SD) at 99.414\% vs. 99.405\%. The impact of Data framing + SD is also evident here as the combination resulted in the lowest false negative of all the models compared.

W.r.t execution time, Fig. 6 shows that the pure ANN was the slowest of all four models, taking over 2 minutes to classify traffic flow. This would be unacceptable in real-time environments, where high-speed data analysis and classification are paramount. In contrast, the variants of ANN based on data framing were significantly faster than both LGR and the pure ANN at just 11s vs 51s and 130s respectively. This shows that breaking traffic into data frames or “windows” and processing them accordingly can be significantly beneficial with regards to processing time.


\setcounter{figure}{6}
\begin{center}
    \includegraphics[scale=0.45]{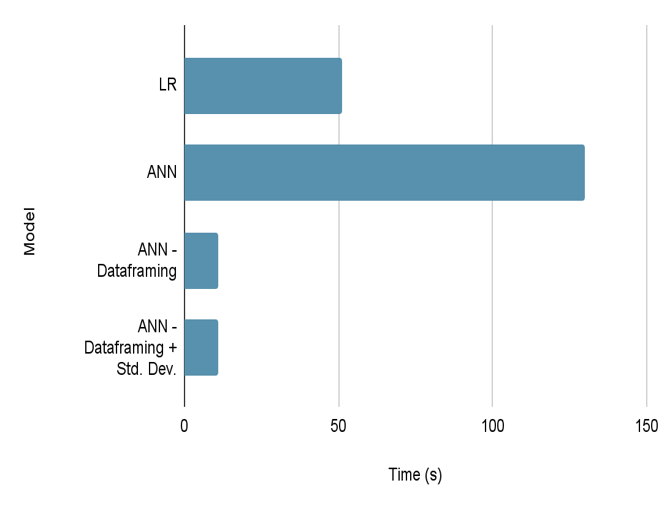}
    \vspace{2mm}
    \parbox[c]{8.3cm}{\footnotesize{Fig.6.} Average Execution Time}
    \label{fig6}
\end{center}

\subsection{Unsupervised Learning}

\setcounter{figure}{7}
\vspace{1mm}
\begin{center}
    \includegraphics[scale=0.4]{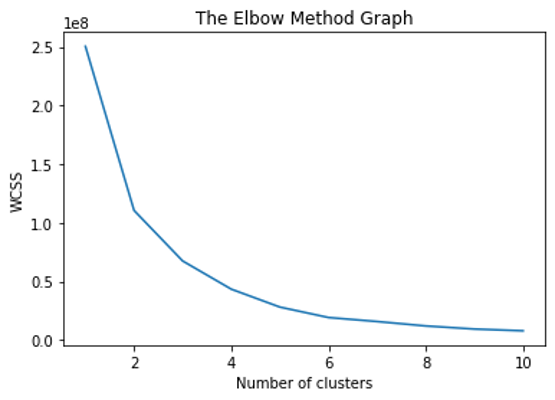}
    \parbox[c]{8.3cm}{\footnotesize{Fig.7.~} K-Means Elbow Method}
    \label{fig7}
\end{center}

Though we knew the number of clusters to expect in the dataset a priori, we still ran the Elbow method \cite{29} to verify this. Fig. 7 shows the result of the Elbow method, with k being re-confirmed to be 2. Running the K-Means classifier with $K=2$, resulted in an accuracy of 96.76\%, with zero false positives.

\subsection{Semi-Supervised Learning}
The labelled outputs from K-Means (unsupervised learning) were used as input to the supervised learning models, in essence creating a form of semi-supervised model. Table \ref{tab2} shows the performance of this hybrid combination. 

\vspace{1mm}
\begin{table*}[!htb]
\centering
\caption{Anomaly Detection Using Semi-Supervised Learning Model}
\label{tab2}
{\footnotesize
\begin{tabular}{l p{2cm} p{2.7cm} p{2.9cm}}
\hline
\textbf{Model} & \textbf{Accuracy} &	\textbf{False Positives} &	\textbf{False Negatives}  \\ \hline 
K-Means + LGR	& 100 & 0 & 0 \\ 
K-Means + ANN & 100 & 0 & 0 \\ 
K-Means+ANN+Dataframing & 99.64 & 0.73 & 0.07 \\ 
K-Means+ ANN+Dataframing+SD. & 99.69 &
0.67 & 0.01 \\ 
\\\hline
\end{tabular}}
\end{table*}

From the table it can be seen that the incorporation of the K-Means classifier resulted in a significant boost in the performance of all the models. Both LGR and the pure ANN resulted in perfect accuracies, zero false positives and zero false negatives.  Similarly, the accuracies of both variants of ANN with data framing increased from 98.842\% to 99.64\% and 99.41 to 99.69\% respectively. Of important note is the reduction in false positive and false negative values of ANN + Dataframing and ANN + Dataframing + SD. respectively. For the former, the false positive dropped from about 2.18\% to just 0.73\%, while the false negative value dropped to 0.07. For ANN + Dataframing + SD, the false positive value dropped to 0.67 \%. The overall improvements in the final results on Table \ref{tab1} compared to Table \ref{tab2} shows the efficacy of our proposed hybrid (semi-supervised) model in detecting malicious attacks. However, the fact that both variants of ANN did not yield 100\% accuracy cannot be ignored. A possible explanation for this is that the dataset was not split into data frames of equal sizes, hence some data frames (especially those at the tail end of the traffic flow) contained less data i.e. less than the window size (12 data points).

\subsection{Statistical Models}
Table \ref{tabStat} summaries the results of the statistical methods used for detecting malicious (abnormal / attack) traffic. For comparison purposes, we also included the result of the pure Logistic Regression model (LGR). 

\vspace{1mm}
\begin{table*}[!htb]
\centering
\caption{Anomaly Detection Using Statistical Models}
\label{tabStat}
{\footnotesize
\begin{tabular}{p{2.5cm} p{2.5cm} p{2.7cm} p{2.9cm}}
\hline
\textbf{Model} & \textbf{Accuracy} &	\textbf{False Positives} &	\textbf{False Negatives}  \\ \hline 
EWMA	& 71.299 & 12.102 & 57.268 \\ 
LDA & 99.837 & 7.797 & 37.870 \\ 
LGR & 99.192 & 0 & 1.6215 \\ \hline
\end{tabular}}
\end{table*}

Compared to LGR, both EWMA and LDA performed poorly w.r.t False Negatives and False Positives. The false positive and negative values in EWMA are understandably high because the model uses simple moving standard deviation and mean of observed samples to determine differentiate attacks. This means that for every subsequent traffic window (120 seconds interval), EWMA would compare the mean and standard deviation of that window with its preceding window. If the difference is much, EWMA flags that window as being attacked. To elaborate, if we assume that little or no data traffic arrive during the first 120 seconds, EWMA establishes a baseline with this first window size using the mean and standard deviation (SD). If during the next few seconds, significant number of legitimate traffic arrive, EWMA calculates the mean and SD of this new block. It then compares the new mean and SD with the baseline. The new values would definitely be higher than the baseline and EWMA would flag this new window as malicious because of the higher traffic count. The reverse is the case with the false negatives.

Being probabilistic (based on Bayes' theorem), LDA expected performs better than EWMA in most of the metrics. However, like EWMA, LDA also struggled with distinguishing between high volume legitimate traffic and malicious traffic. This problem becomes more pronounced when low traffic windows(s) is/are followed by window(s) with slightly higher traffic counts. In such instances, the succeeding window(s) would be classified as malicious even if there are not.
 
\subsection{Prediction}
As stated earlier, three prediction models were considered and their results are summarized on Table \ref{tabPred}. Of the three models compared, LGR performed the best, with a prediction accuracy of 98.6\%. It was closely followed by KRR at approx. 98\%. SVR was the least accurate of the lot at 94.64\%. For $R^2$, values closer to 1 are desirable, and depicts the “closeness” of predicted values to the actual values. For the three models, the same trend is observed with $R^2$ scores, as LGR led with a score of approx. 0.94, followed by KRR at 0.91. SVR scored 0.76, implying that its prediction curve differed greatly from the actual curve. Finally, for RMSE, values closer to 0 are desirable as they indicate lower prediction errors. Once again, LGR was the least error prone as it had the lowest RMSE values, followed by KRR with a score of 0.1439. However, both models were less error prone than SVR with a RMSE value of 0.2314. We can thus conclude that LGR is the best predictor, while KRR is a close alternative. With such high RMSE value, SVR is a less than ideal predictor in our use case.

\vspace{1mm}
\begin{table*}[!htb]
\centering
\caption{Comparison of Results of the Prediction Models}
\label{tabPred}
{\footnotesize
\begin{tabular}{p{2cm} p{3cm} p{3cm} p{3cm}} \\
\hline
\textbf{Metric} & \textbf{KRR} & \textbf{LGR} &	\textbf{SVR} \\ \hline
Accuracy  & 97.93\% & 98.60\% & 94.64\% \\ 
$R^2$  & 0.9054 & 0.9361 & 0.7555  \\ 
RMSE  & 0.1439 & 0.1183 & 0.2314 \\ \hline
\end{tabular}}
\end{table*}

To visualize these results we plotted graphs of actual values versus those predicted by the prediction models (KRR, LGR and SVR). For each model we took snapshots showing its predictions for the next 15 minutes (900 seconds). Furthermore, to show the scalability of the predictive models, we also took snapshots of predictions for the next 2 to 4 hours. These snapshots are depicted in Figs. \ref{fig8} to \ref{fig13}, where the blue lines represent predicted values, while the red lines indicate actual values.

\setcounter{figure}{8}
\begin{figure*}[!htb]
\centering
\subfigure[KRR predictions for the next 15 minutes.]{
\includegraphics[scale=0.4]{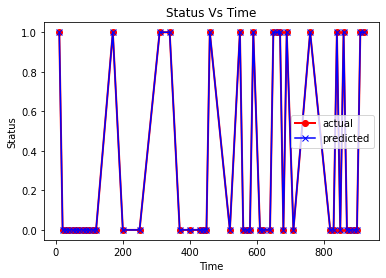} 
\label{fig8}
}
\subfigure[LGR predictions for the next 15 minutes.]{
\includegraphics[scale=0.4]{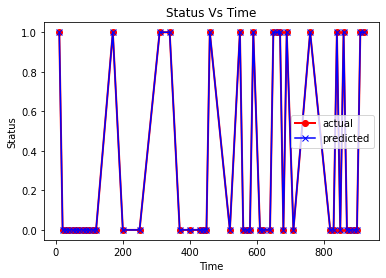} 
\label{fig9}
}
\subfigure[SVR predictions for the next 15 minutes.]{
\includegraphics[scale=0.4]{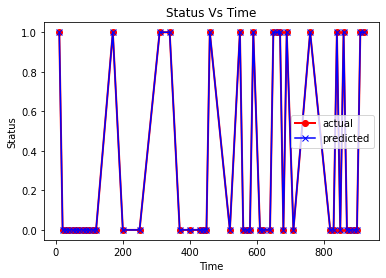} 
\label{fig10}
}\\[-1.5mm]
\caption{Comparison of predictions for the next 15 minutes.} 
\label{fig8-10}
\end{figure*}

\setcounter{figure}{9}
\begin{figure*}[!htb]
\centering
\subfigure[KRR predictions for the next 3 to 4 hours.]{
\includegraphics[scale=0.4]{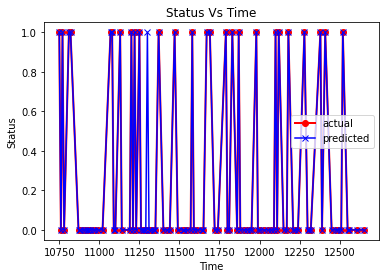} 
\label{fig11}
}
\subfigure[LGR predictions for the next 3 to 4 hours.]{
\includegraphics[scale=0.4]{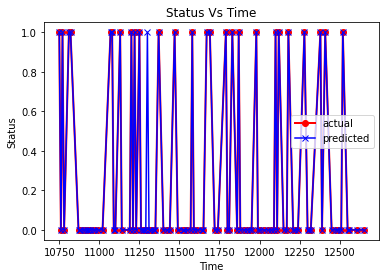} 
\label{fig12}
}
\subfigure[SVR predictions for the next 3 to 4 hours.]{
\includegraphics[scale=0.4]{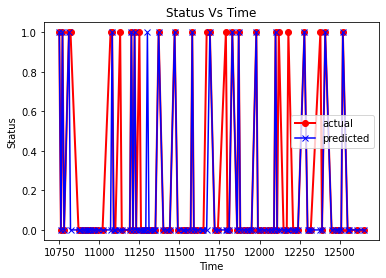} 
\label{fig13}
}\\[-1.5mm]
\caption{Comparison of predictions for the next 3 to 4 hours.} 
\label{fig11-13}
\end{figure*}

Figs. \ref{fig8} to \ref{fig10} show plots of status versus time for KRR, LGR and SVR respectively. Status values are binary and can only be 0 or 1, where 0 means no attack is predicted and 1 means that an attack might take place. For all three models similar graphs were observed for the 15 minutes time frame. Figs. \ref{fig11} to \ref{fig13} show the predictions of the three models for the next 3 to 4 hours. KRR and LGR had similar graphs, with both models coincidentally wrongly predicting the occurrence of an attack at around the $11,260^{th}$ second marks. The results of SVR's predictions for the same period are shown in Fig. \ref{fig13}, with SVR making wrong predictions on 10 different occasions. Interestingly, like KRR and LGR, SVR also wrongly predicted an imminent attack at the same $11,260^{th}$  second mark. 

Overall these results show that LGR and KRR are better prediction models than SVR for our use case. With attack prediction accuracies of approximately 98\% for both LGR and KRR models, it can be inferred that regression models can be used to predict potential DDoS attacks in IoT networks. For both LGR and KRR, the inaccurate predictions were in instances where they assumed that attacks would occur when none occurred. These wrong predictions or false alarms, though leading to unnecessary deployment of defensive mechanisms, are preferable to the reverse case. In the reverse case, as observed with SVR, the model gives a false sense of security by predicting that no attack would occur, when imminent threats abound. We therefore consider the wrong predictions of KRR and LGR as “erring on the side of caution”.

\section{Conclusion} \label{s6}
In this study, the accuracy and timeliness of supervised, unsupervised, and semi-supervised machine learning techniques for detecting Distributed Denial of Service (DDoS) attacks in Cyber Physical-Internet of Things Systems (CPS-IoT) were explored. CPS-IoT systems often rely on two well known protocols for data transmission, namely HTTP and MQTT, both of which are built upon TCP/IP, hence vulnerable to TCP/IP targeted attacks. DDoS attacks are common to TCP/IP, thus pose a potential threat to the security, dependability and safety of CPS-IoT systems. In this work, five machine learning models (ML) and two statistical models were considered for modelling DDoS attacks in IoT networks (TCP/IP-based). These are Logistic Regression (LGR), Artificial Neural Networks (ANN), K-Means, Kernel Ridge Regression (KRR),  Support Vector Regression (SVR), Exponentially Weighted Moving Average (EWMA) and the Linear Discriminant Analysis (LDA).

In distinguishing between normal traffic and bogus (attack) traffic, two supervised ML classifiers were used - LGR and ANN (and two slight variations of ANN based on slicing). LGR gave a classification accuracy of 99.19\%, a false positive rate of 1.62\%, and an average detection latency of 51 seconds from the initiation of the attack. The ANN model, on the other hand, had better accuracy at 99.41\% and lower false negative value of 0.67 \%, but was extremely slow at 130 seconds. We introduced slicing, and split the traffic into fixed windows sizes, before applying ANN. This slicing improved the false negative values and significantly cut down the detection time to just 11 seconds. We then considered the K-Means unsupervised ML model, which resulted in 96.76\% classification accuracy. Finally, we developed semi-supervised ML models by combining the K-Means with the ANN and LGR. These combinations resulted in a classification (detection) accuracy of 100\% with near zero false positives across all models. Compared to the ML models, the statistical models performed poorly w.r.t false positive and negatives. 

We also examined the use of regression models to support network administrators in transiting from reactive to proactive network management approach. LGR, KRR, and SVR were investigated for their abilities to correctly predict attacks before they occur. LGR gave the best prediction accuracy at 98.6\%, followed by KRR at 97.9\%, while SVR had the worst performance at 94.64\%. The $R^2$ values for the LGR and KRR were 0.94 and 0.91 respectively, representing closeness to actual values; while their RMSE values were respectively 0.12 and 0.14. SVR was significantly off the mark for these metrics. In essence, LGR and KRR are both capable for predicting imminent threats, with LGR being slightly better. 

This work only considered traffic counts, time and status in determining DDoS attack. In future works, other features such as source and destination IP addresses or ports can be considered. Similarly, only the possible future attack times were considered, potential future research could consider the target machine or subnet. Finally, in CPS-IoT systems, attack detection and prediction through the use of network topology graphs could be another avenue for future research work.

\end{document}